\documentclass[aps,pre,longbibliography,showkeys,groupedaddress,amsmath,amssymb,floatfix,twocolumn]{revtex4-1}

\usepackage{color} 
\usepackage{todonotes}
\usepackage{verbatim}

\usepackage{lmodern}						

\begin{document}


\title{On the onset of synchronization of Kuramoto oscillators in scale-free networks}
\author{Thomas Peron$^{1,4}$}
\email{thomaskaue@gmail.com}
\author{Bruno Messias$^{2}$}
\author{Ang\'elica S. Mata$^{3}$}
\author{Francisco A. Rodrigues$^{1}$}
\author{Yamir Moreno$^{4,5,6}$}
\affiliation{$^{1}$Institute of Mathematics and Computer Science, University of S\~ao Paulo,
S\~ao Carlos, S\~ao Paulo, Brazil}
\affiliation{$^{2}$S\~ao Carlos Institute of Physics, University of S\~ao Paulo,
S\~ao Carlos, S\~ao Paulo, Brazil}
\affiliation{$^{3}$Departamento de F\'isica, Universidade Federal de Lavras, 37200-000 Lavras, MG, Brazil}
\affiliation{$^{4}$Institute for Biocomputation and Physics of Complex Systems (BIFI), University of Zaragoza, Zaragoza 50018, Spain}
\affiliation{$^{5}$Department of Theoretical Physics, University of Zaragoza, Zaragoza 50009, Spain}
\affiliation{$^{6}$ISI Foundation, Torino, Italy}

\begin{abstract}
Despite the great attention devoted to the study of phase oscillators on complex networks in the last two decades, 
it remains unclear whether scale-free networks exhibit a nonzero critical coupling strength for the onset of synchronization in the thermodynamic limit. Here, we systematically compare predictions from the heterogeneous degree mean-field (HMF) and the quenched mean-field (QMF) approaches to extensive numerical simulations on large networks. 
We provide compelling evidence that the critical coupling vanishes as the number of oscillators 
increases for scale-free networks characterized by a power-law degree distribution with an exponent $2 < \gamma \leq 3$, in line with what has been observed for other dynamical processes in such networks. For $\gamma > 3$, we show that the critical coupling remains finite, in agreement with HMF calculations and highlight phenomenological differences between critical properties of phase oscillators and epidemic models on scale-free networks. Finally, we also discuss at length a key choice when studying synchronization phenomena in complex networks, namely, how to normalize the coupling between oscillators. 
\end{abstract}

\pacs{05.40.-a, 05.45.Xt, 87.10.Ca}
\maketitle 

\section{Introduction}

Synchronization processes are pervasively observed in 
a wide range of physical, chemical, technological and 
biological systems~\cite{pikovsky2003synchronization}. These phenomena
can, to a great extent, be described by models of coupled phase oscillators. Arguably,
one of the most studied models in this context is the one proposed by Kuramoto~\cite{acebron2005kuramoto}, which  
in the last decade was
extensively investigated when the oscillators are placed on complex networks (see~\cite{arenas2008synchronization,rodrigues2016kuramoto} and
references therein). A key
question addressed in these studies is how the heterogeneous 
connectivity pattern impacts on the onset of synchronization -- or, in other words,  
how the critical coupling strength required for the emergence of collective 
motion is affected by the network topology. 

The relationship between structure 
and synchronous dynamics has been studied in many scenarios: from 
homogeneous and unclustered networks to heterogeneous and modular ones, in addition to variations of phase oscillator models including correlations between intrinsic dynamics and local topology~\cite{arenas2008synchronization,rodrigues2016kuramoto}. Yet, despite the notorious 
advances achieved over the past years, fundamental questions regarding 
the collective dynamics of large ensembles of oscillators still remain elusive. 
One of these problems is whether the critical coupling strength 
for the onset of synchronization remains finite in the thermodynamic
limit for scale-free (SF) networks characterized by a power-law degree distribution with an exponent $2<\gamma \le 3$. Another important question concerns the very definition of the coupling strength in the dynamical equations. This paper will address both challenges.

The above questions were already pointed out in the first work that dealt with the dynamics of Kuramoto oscillators on heterogeneous scale-free structures~\cite{moreno2004synchronization}. There, the authors remarked on the supposed finite magnitude of the critical coupling 
and highlighted the apparent contrast of the Kuramoto dynamics with
epidemic spreading and percolation -- processes which were already known to 
exhibit vanishing critical points in the thermodynamic limit for SF topologies. 
Subsequent theoretical approaches~\cite{ichinomiya2004frequency,lee2005synchronization} estimated via mean-field approximations 
that, in the absence of degree-degree correlations, the critical coupling should converge to zero
as the number of oscillators tends to infinity -- similarly to what happens for other dynamical processes on networks~\cite{boccaletti2006complex}. However, later investigations reported significant 
deviations between predictions of mean-field theories and
numerical simulations~\cite{restrepo2005onset}, 
casting further doubts on the validity of the classical result  
on the nonexistence of a synchronization threshold~\cite{rodrigues2016kuramoto,arenas2008synchronization,dorogovtsev2010lectures}. 
        
One clear difficulty for a precise estimation of 
the onset of synchronous motion is, naturally, the sizes of the simulated networks. 
Indeed, the first hypotheses on the existence or absence of a critical point
in the Kuramoto dynamics were supported by numerical experiments considering 
populations with sizes of the order of up to $10^4$ oscillators~\cite{moreno2004synchronization,ichinomiya2004frequency,lee2005synchronization,restrepo2005onset} -- a value that potentially limits the accuracy of finite-size analysis and calculations, especially in what concerns 
the detection of the onset of synchronization for highly heterogeneous structures. It is noteworthy to mention, though, that recent contributions (see, e.g.~\cite{rodrigues2016kuramoto,hong2007finite,hong2013link,um2014nature}) have investigated
finite-size effects of the dynamics and reported excellent agreement between simulation and mean-field theories. However, most of those analyses have
focused on Erd\H{o}s-R\`enyi (ER) random graphs and SF networks with degree exponent $\gamma > 3$, situations 
in which the critical coupling is expected to be finite (according to the heterogeneous degree mean-field approximations)~\cite{rodrigues2016kuramoto}. Of particular interest is a recent contribution~\cite{juhasz2019critical}, where the authors investigated finite-size effects of ER graphs, reaching networks with very large sizes (up to $N = 2^{27}$ nodes).

Another possible source of disagreement between mean-field theories and numerical simulations in the estimation 
of the critical coupling strength is the consideration of different definitions of order parameters~\cite{arenas2008synchronization,boccaletti2006complex}. Very recently, Yook and Kim~\cite{yook2018two} performed a thorough comparison between the classical Kuramoto order parameter and the order parameter accounting for heterogeneous degree distributions. The authors verified that, indeed, the definition of the order parameter crucially affects
the assessment of the asynchronous state in highly heterogeneous SF networks. 
However, although simulations with networks of size up to $10^7$ oscillators were carried out, 
it is not clear from~\cite{yook2018two} how the transition point behaves as the network size increases.
Therefore, the question regarding whether or not there is a well defined critical 
coupling for the onset of synchronization in SF networks with 
$2 < \gamma \leq 3$ has remained without a concluding answer.        
        
In order to address this problem, and given the difficulties in performing very large numerical simulations, here we adopt an alternative approach: we perform a systematic comparison between simulations 
and the results derived using the heterogeneous degree mean-field (HMF) and 
the quenched mean-field (QMF) formulations in networks of sizes up to $N = 3\times10^6$ nodes. We show that the critical coupling predicted by both the HMF and the QMF agrees with the values measured in numerical experiments for networks with power-law exponent $\gamma \leq 3$, hence providing stronger evidence that the critical coupling of such systems vanishes in the thermodynamic limit. For SF networks whose degree distribution has a finite second statistical moment, we find that the onset of synchronization remains constant in the thermodynamic limit. Furthermore, we highlight differences between the critical behavior of synchronization dynamics and that found in the disease spreading process. In particular, we verify that HMF correctly predicts a finite critical threshold in the thermodynamic limit for $\gamma > 3$, in contrast to results obtained in the context of epidemic dynamics~\cite{ferreira2012epidemic,mata2013pair}. 

Additionally, we also revisit another issue debated in early works on phase oscillator models on networks: how to define and properly normalize the coupling function in the dynamical equations. In particular, we verify that some choices previously considered as appropriate for SF networks actually induce undesired dependences on the system's size, including the increase of the onset of synchronization as networks become larger, and an infinite coupling strength that locks low degree nodes in the thermodynamic limit of highly heterogeneous networks. The rest of the paper is organized as follows: In Sec.~\ref{sec:MF}, we provide a brief review of
the mean-field approximations to treat coupled oscillators in heterogeneous networks. In Sec.~\ref{sec:Kc},
we compare the estimations by mean-field theories with numerical simulations. Section~\ref{sec:norm}
is devoted to the discussion on the coupling normalization. We give our conclusions in Sec.~\ref{sec:conclusion}.

\section{Mean-field theories for Phase oscillators in heterogeneous networks}
\label{sec:MF}
 
In this section, we provide a brief review of the main analytical approximations used to deal with ensembles of phase oscillators in heterogeneous networks. The Kuramoto model consists of the following system of equations~\cite{arenas2008synchronization,rodrigues2016kuramoto}
\begin{equation}
\dot{\theta}_i(t) = \omega_i + K \sum_{j=1}^N A_{ij} \sin(\theta_j - \theta_i),
\label{eq:KM_net}
\end{equation}
where $\theta_i$ and $\omega_i$ are the phase and natural frequency of the $i$-th oscillator, respectively; $K$ is the coupling strength, and $\mathbf{A}$ is the adjacency matrix, with $A_{ij} = 1$ if nodes $i$ and $j$ are connected, and $0$ otherwise. 

In order to assess the overall synchrony of an ensemble of oscillators, Kuramoto~\cite{acebron2005kuramoto}
introduced the order parameter
\begin{equation}
R e^{\mathrm i\psi(t)} = \frac{1}{N} \sum_{j=1}^N e^{\mathrm i\theta_j(t)},
\label{eq:order_parameter_Kuramoto}
\end{equation}
where $R$ and $\psi$ are the magnitude and phase of the centroid associated to the $N$ points $e^{ \mathrm{i} \theta_j(t)}$ in the complex plane, respectively. If phases are uniformly distributed 
over $[0, 2\pi]$, it follows that $R \approx 0$, whereas $R \approx 1$ if oscillators 
rotate grouped into a synchronous cluster.

In contrast to the case of globally connected populations, the original analytical treatment via a self-consistent analysis by Kuramoto~\cite{acebron2005kuramoto} cannot be directly extended to the network case. The reason for this relies in the fact
that Eqs.~\ref{eq:KM_net} are not exactly decoupled by a global order parameter. Instead, an exact decoupling is only achieved by defining
local order parameters as 
\begin{equation}
r_i e^{\mathrm i\psi_i(t)}= \sum_{j=1}^N A_{ij} e^{\mathrm i\theta_j(t)},
\label{eq:local_parameter}
\end{equation}
which leads to 
\begin{equation}
\dot{\theta}_i(t) = \omega_i + Kr_i \sin(\psi_i - \theta_i).
\label{eq:decoupled_local_order_parameter}
\end{equation}

In this paper, we consider the oscillators frequencies $\omega_i$ to be distributed 
according to a smooth and unimodal distribution $g(\omega)$ centered at $\omega=0$. 
By inserting the fixed point solution $(\dot{\theta}_i(t)=0)$ of the equation above into Eq.~\ref{eq:local_parameter}, and performing a self-consistent analysis of the resulting equation, one arrives at the critical coupling given by~\cite{restrepo2005onset,rodrigues2016kuramoto}
\begin{equation}
K_c^{\rm{QMF}} = \frac{2}{\pi g(0)} \frac{1}{\Lambda_{\max}}, 
\label{eq:KcQMF}
\end{equation}
where $\Lambda_{\max}$ is the largest eigenvalue of $\mathbf{A}$. The latter result
was first derived in~\cite{restrepo2005onset}, with what the authors called \textit{perturbation theory} of the Kuramoto model on complex networks. Henceforth, we refer to Eq.~\ref{eq:KcQMF} as the QMF critical coupling strength, owing to the
similarity with epidemic thresholds derived with techniques that preserve the quenched structure of the network~\cite{castellano2010thresholds}. To gain further insights on the predictions of Eq.~\ref{eq:KcQMF} to the dynamics on SF networks, we recall the result~\cite{chung2004spectra}
\begin{equation}
\Lambda_{\max}\sim\begin{cases}
\frac{\langle k^{2}\rangle}{\langle k\rangle} & \mbox{ if }\frac{\langle k^{2}\rangle}{\langle k\rangle}>\sqrt{k_{\max}}\ln(N),\\
\sqrt{k_{\max}} & \mbox{ if }\sqrt{k_{\max}}>\frac{\langle k^{2}\rangle}{\langle k\rangle}\ln^{2}(N),
\end{cases}
\label{eq:chung_result}
\end{equation}
where $k_{\max}$ is the maximum degree or the network. In uncorrelated
SF networks, $k_{\max}$ scales as $k_{\max} \sim N^{1/2}$ if $2 < \gamma \leq 3$, and
$k_{\max} \sim N^{1/(\gamma - 1)}$, for $\gamma > 3$. By noticing further that
$\langle k^2 \rangle/ \langle k \rangle \sim k_{\max}^{3-\gamma} \ll \sqrt{k_{\max}}$, we then estimate~\cite{castellano2010thresholds}
\begin{equation}
K_{c}^{{\rm QMF}}\simeq\frac{2}{\pi g(0)}\times\begin{cases}
\frac{\langle k\rangle}{\langle k^{2}\rangle} & \mbox{ if }2<\gamma<5/2,\\
\frac{1}{\sqrt{k_{\max}}} & \mbox{ if }\gamma>5/2.
\end{cases}
\end{equation}
Therefore, according to the QMF approach, the critical coupling $K_c$ should vanish in the thermodynamic limit as $k_{\max}$ diverges, even if $\langle k^2 \rangle$ remains finite (i.e., the case when $\gamma>3$).

Another way of modeling synchronization processes on networks is by virtue of the annealed 
network approximation~\cite{rodrigues2016kuramoto}. It consists of replacing the elements of the adjacency matrix $A_{ij}$ by its ensemble average $\tilde{A}_{ij}$, which corresponds to the probability
that two nodes, $i$ and $j$, are connected in the configuration model; that is,   
\begin{equation}
\tilde{A}_{ij} = \frac{k_i k_j}{N \langle k \rangle },
\label{eq:annealed_networks}
\end{equation}
where $k_i$ is the degree of node $i$. Substituting Eq.~\ref{eq:annealed_networks} into Eq.~\ref{eq:KM_net} yields
\begin{equation}
\dot\theta_i(t) = \omega_i + \frac{K k_i}{N \langle k \rangle} \sum_j k_j \sin(\theta_j - \theta_i). 
\label{eq:eq_motion_HMF}
\end{equation} 
The previous equation motivates the definition of the following order parameter
\begin{equation}
r e^{\mathrm i\psi(t)} = \frac{1}{N\langle k \rangle} \sum_{j=1}^N k_j e^{\mathrm i \theta_j(t)}.
\label{eq:order_parameter_HMF}
\end{equation}
\begin{figure}[!t]
	\centering
	\includegraphics[width=0.9\columnwidth]{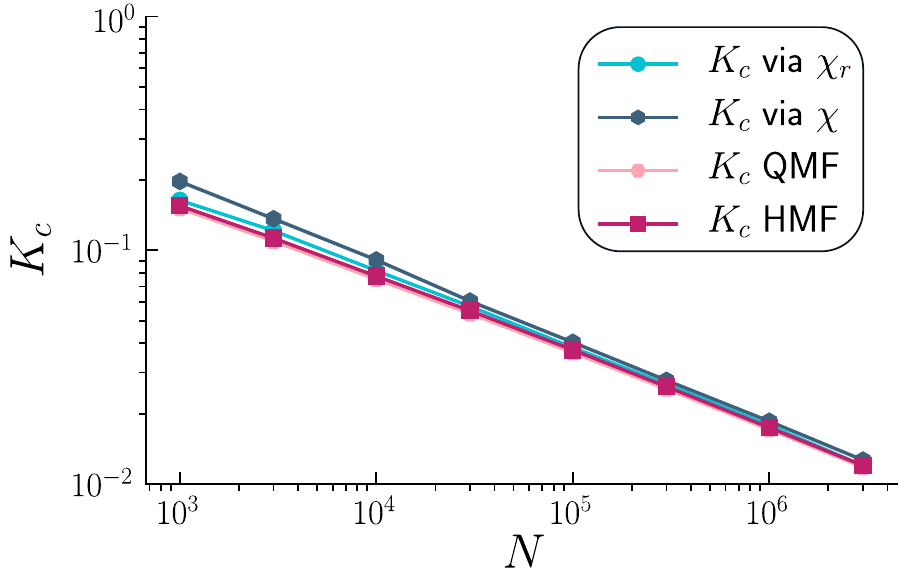}
	\caption{(Color online) Comparison between the estimations of $K_c$ by susceptibilities $\chi$ (Eq.~\ref{eq:chi}) and $\chi_r$ (Eq.~\ref{eq:chi_r}). Networks generated according to the UCM with degree distribution $P(k) \sim k^{-\gamma}$, with $\gamma = 2.25$ and $k_{\min} = 5$. Natural frequencies are assigned according to Eq.~\ref{eq:frequency_assignment_II}. Each point is an average over 100 network realizations. Error bars are smaller than symbols.}%
	\label{fig:bothChis}%
\end{figure}
Equation~\ref{eq:annealed_networks} is equivalent to the so-called heterogeneous degree mean-field approximation (HMF)~\cite{rodrigues2016kuramoto} and leads to the definition of the order parameter 
in Eq.~\ref{eq:order_parameter_HMF}. Essentially, in the HMF approximation, one assumes
that the network topology (initially fully represented by the adjacency matrix $\mathbf{A}$) 
is abstracted in the degree distribution $P(k)$; that is, nodes are coarse-grained according to their degrees and the oscillators become statistically equivalent, differing only by the parameters $k_i$ and $\omega_i$.  
\begin{figure*}
	\centering
	\includegraphics[width=0.8\textwidth]{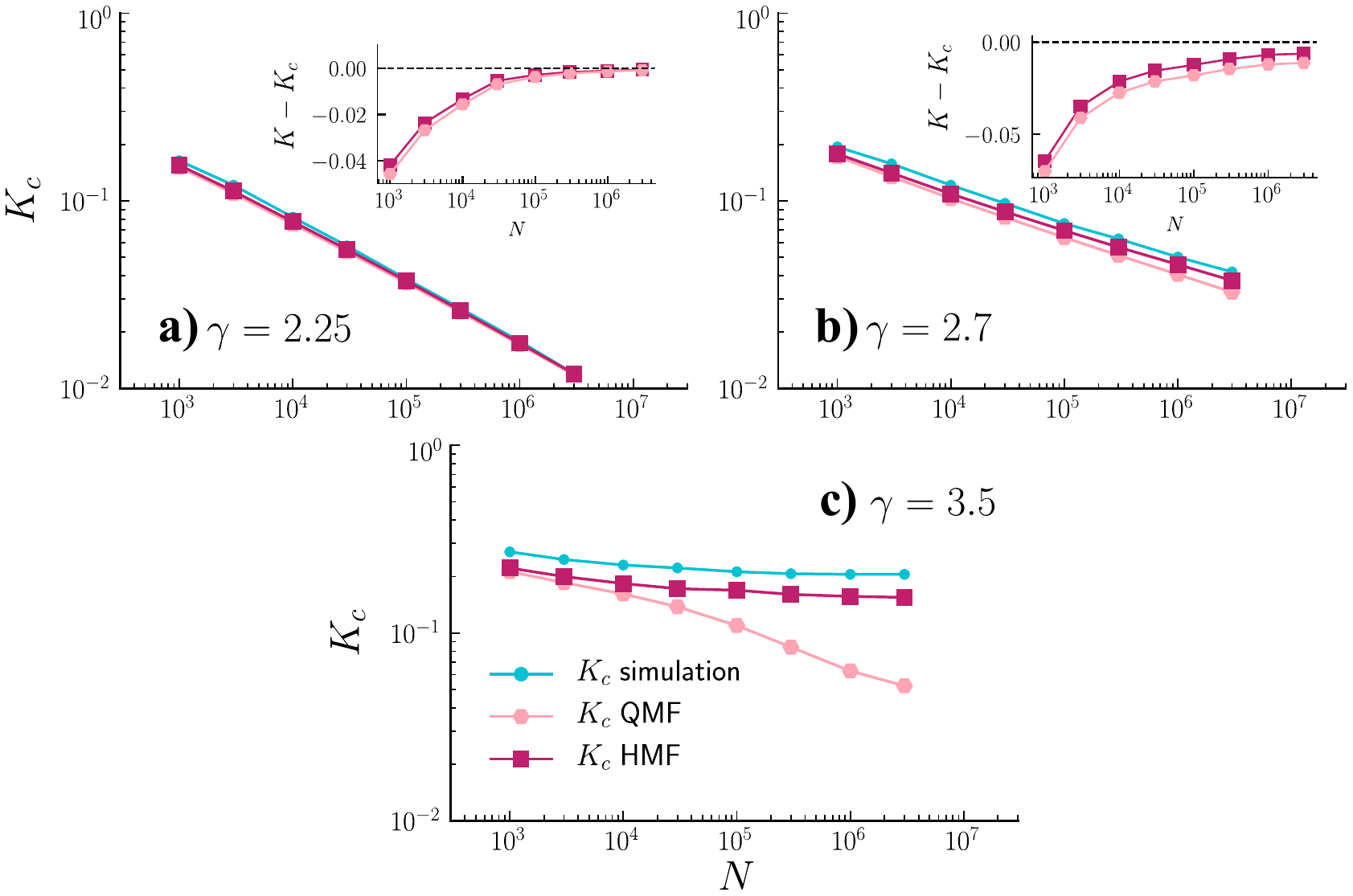}
	\caption{(Color online) Critical coupling $K_c$ against network size $N$ for UCM networks wtih power-law exponent (a) $\gamma = 2.25$, (b) $\gamma = 2.7$, (c) $\gamma = 3.5$. All networks have $k_{\min} = 5$. Insets in (a) and (b) depict the difference between numerical estimation of $K_c$ and mean-field theories. Each point is an average over 100 network realizations. Error bars are smaller than symbols.}%
	\label{fig:KcvsN}%
\end{figure*}

By decoupling Eqs.~\ref{eq:eq_motion_HMF} with Eq.~\ref{eq:order_parameter_HMF}, and 
performing a self-consistent analysis of the equations, one can show that the onset 
of synchronization within the annealed approximation occurs when~\cite{rodrigues2016kuramoto,ichinomiya2004frequency}   
\begin{equation}
K>K_c^{\rm{HMF}} = \frac{2}{\pi g(0)} \frac{\langle k \rangle}{\langle k^2 \rangle}, 
\label{eq:KcHMF}
\end{equation}
where $\langle k^n \rangle = \sum_k k^n P(k)$ is the $n-$th moment of the degree distribution $P(k)$. 
   
As previously mentioned, it has been recently shown~\cite{yook2018two} that the traditional order parameter (Eq.~\ref{eq:order_parameter_Kuramoto}) and the one introduced by the HMF approximation yield different
results when assessing the synchronization of networks. In particular, the latter 
overestimates the level of coherence among the oscillators in the asynchronous 
regime for SF networks. This effect is particularly evident in networks having hubs whose 
degree scales with $\mathcal{O}(N)$; however, discrepancies between $R$ and $r$ are also likely to emerge for networks with power-law exponent $\gamma >3$~\cite{yook2018two}. Therefore, in this paper, we evaluate the onset synchronization numerically using the standard order parameter $R$ in Eq.~\ref{eq:order_parameter_Kuramoto}. 

Our goal is to systematically investigate the behavior of the onset of synchronization as the size of SF networks increases, comparing 
the theoretical predictions provided by the current mean-field approaches. Seeking to keep 
the source of fluctuations across network realizations to a minimum, we assign 
natural frequencies deterministically according to~\cite{hong2013link}
\begin{equation}
\frac{i}{N}-\frac{1}{2N}=\int_{-\infty}^{\omega_{i}}g(\omega)d\omega,
\label{eq:frequency_assignment_I}
\end{equation}
which for the Lorentzian distribution $g(\omega) = \frac{\Delta}{\pi} \frac{1}{\omega^2 + \Delta^2}$ 
yields
\begin{equation}
\omega_{i}=\Delta\tan\left[\frac{i\pi}{N}-\frac{(N+1)\pi}{2N}\right],\;i=1,\dots,N.
\label{eq:frequency_assignment_II}
\end{equation} 
In this way, we generate a set of quasi-uniformly spaced frequencies, removing, thus, the disorder introduced by different realizations of frequencies~\cite{hong2013link}. 

\section{Critical coupling of uncorrelated scale-free networks}
\label{sec:Kc}

All networks analyzed in this section were generated following the 
uncorrelated configuration model (UCM)~\cite{catanzaro2005generation}
considering a power-law degree distribution $P(k) \sim k^{-\gamma}$ with the cutoff $k_{\max} \sim N^{1/2}$ for $\gamma \leq 3$, and 
$k_{\max} \sim N^{1/(\gamma -1)}$, for $\gamma > 3$. Furthermore, in order 
to avoid sample-sample fluctuations on $k_{\max}$, for each value of $N$, 
we fixed $k_{\max} = \langle k_{\max} \rangle$ across the network realizations.

Typically, the critical coupling strength of finite networks can be estimated numerically 
via detecting the divergent peak of the susceptibility 
\begin{equation}
\chi = N (\langle R^2 \rangle_t - \langle R \rangle_t^2), 
\label{eq:chi}
\end{equation}
where $\langle \cdots \rangle_t$ denotes a temporal average. However, we employ 
the modified susceptibility defined as~\cite{ferreira2012epidemic}
\begin{equation}
\chi_r = N \frac{(\left\langle R^2 \right\rangle_t - \left\langle R \right\rangle_t^2)}{\langle R \rangle_t}, 
\label{eq:chi_r}
\end{equation}
As with the definition in Eq.~\ref{eq:chi}, the modified susceptibility also exhibits a peak at $K=K_c$. Nonetheless, analogous forms of $\chi_r$ have been shown to be better suited to detect transition points in epidemic spreading and contact processes in networks with diverging $\langle k^2 \rangle$~\cite{ferreira2012epidemic,mata2015multiple,mata2014heterogeneous}. Thus, motivated by those results, we extend this measure for the detection of onset of the synchronous state. Our choice is confirmed by the numerical results presented in Fig.~\ref{fig:bothChis}. For $\gamma = 2.25$, the critical points estimated via Eq.~\ref{eq:chi_r} are in better agreement with HMF and QMF theories than that estimated via Eq.~\ref{eq:chi}, especially for low values of $N$. Similar results are found for different values of $\gamma$. Thus, we henceforth detect the critical points via $\chi_r$. 

Let us now analyze how the mean-field theories perform in comparison 
with simulations for the different regimes of $\gamma$. First, for $\gamma < 5/2$, as
discussed in the previous section, both HMF and QMF predict a vanishing $K_c$, 
which should scale with $\langle k \rangle / \langle k^2 \rangle$. Indeed, as it is seen in Fig.~\ref{fig:KcvsN}(a), for $\gamma = 2.25$, both theories predict quite accurately the onset of synchronization.  

Discrepancies between the approximations appear when $\gamma > 5/2$. To be precise, in this regime, HMF
yields $K_c \sim \langle k \rangle /\langle k^2 \rangle$, while QMF gives $K_c \sim k_{\max}^{-1/2}$. As depicted in Fig.~\ref{fig:KcvsN}(b), the mean-field theories provide a satisfactory approximation of the synchronization thresholds for networks with $\gamma = 2.7$. Note that, 
although QMF contains in its formulation the whole information about the network topology, it performs slightly worse than HMF (see inset). Similar dependences with the system size are found for epidemic thresholds in SF networks with $5/2 < \gamma < 3$~\cite{mata2013pair,ferreira2012epidemic}. 


For $\gamma = 3.5$ (Fig.~\ref{fig:KcvsN}(c)), we observe that the numerical calculation of 
$K_c$ converges to a constant value as $N$ increases, in agreement with the HMF prediction, whereas 
QMF theory clearly fails in capturing the onset of synchronization. 
That is, while simulations show that large SF networks in this case exhibit a finite synchronization
threshold, QMF reveals a vanishing $K_c$. Furthermore, it is interesting to point out discrepancies between 
synchronization and the epidemic spreading described by the susceptible-infected-susceptible (SIS)
model \cite{harris74,epidemics} regarding the dependence on the system size for $\gamma > 3$. 
In contrast to the finite onset of synchronization seen in Fig.~\ref{fig:KcvsN}(c), epidemic thresholds of the SIS model
are known to decay as $N$ increases for $\gamma >3$~\cite{mata2013pair,ferreira2012epidemic}. In fact, Chatterjee and Durrett
\cite{Chatterjee09} proved rigorously that, for uncorrelated random networks with a power-law degree 
distribution $P(k) \sim k^{-\gamma}$ with any $\gamma$, the SIS model presents an unstable absorbing phase in the thermodynamic limit, resulting in a null epidemic threshold. Afterwards,
Bogu\~n\'a {\it et al.} \cite{boguna2013nature} physically interpreted this proof with a 
semi-analytical approach taking into account a long-range reinfection mechanism between hubs and found a vanishing epidemic threshold including for $\gamma>3$.

Actually, the behavior of the SIS model is distinct and more intricate than other dynamical
processes that also present a phase transition from active to inactive states. This epidemic model is governed by mutual 
activation of hubs.  Outliers, a small amount of vertices with connectivity much larger than the other nodes of the 
network, can sustain localized epidemics for long times. This phenomenon causes a double-peaked shape in the susceptibility
curve~\cite{mata2013pair,ferreira2012epidemic} and the emergence of Griffiths effects~\cite{Cota}--in this case, QMF captures the
peak associated to the activation of the largest hub in the network~\cite{mata2015multiple}. 
Surprisingly, simulations with networks with $N = 10^7$ (not shown here) did not reveal signs of
multiple peaks in susceptibility curves of Kuramoto oscillators. However, with the aim of understanding 
the nature of the threshold in epidemic models on uncorrelated random networks, recent works \cite{mata2014heterogeneous, sander, Cota2018}
showed that different epidemic models such as, for instance, susceptible-infected-recovered-susceptibility \cite{anderson}, 
contact process~\cite{cp}, the generalized SIS model with weighted infection rates \cite{KJI} and 
other alterations of the SIS model \cite{Cota2018}, have a finite threshold in the thermodynamic limit. 
This behavior is related to standard phase transitions given by collective activation processes
involving essentially the whole network, as observed in the synchronization phenomenon of the Kuramoto oscillators.

At last, the results in Fig.~\ref{fig:KcvsN} point to a different scenario as the one in~\cite{restrepo2005onset} 
regarding the accuracy of mean-field theories. More precisely, in Fig.~\ref{fig:KcvsN} we see that 
HMF exhibits an excellent agreement with numerical simulations for $\gamma = 2.25$ and $2.7$, and a qualitative agreement with the scaling with $N$ for $\gamma = 3.5$. Conversely, Ref.~\cite{restrepo2005onset} found that HMF agrees best with the numerical results obtained for $\gamma > 3$, while significantly deviating from simulations for $2 < \gamma < 5/2 $; i.e., the opposite situation 
observed in Fig.~\ref{fig:KcvsN}. These discrepancies are possibly due to 
structural correlations induced by the large artificial cutoffs ($k_{\max} \sim N$) and the
relative small size of the SF networks ($N \sim 10^3$) considered in~\cite{restrepo2005onset}. 

\section{Coupling normalization}
\label{sec:norm}

The size dependence of the onset of synchronization on the system's size brings back to attention a topic intensively debated in early studies of network synchronization~\cite{arenas2008synchronization,rodrigues2016kuramoto}, namely, the choice for the normalization 
of the coupling function. When dealing with phase oscillators on networks, it is a
common practice to let the oscillators interact through unnormalized couplings, as in Eq.~\ref{eq:KM_net}. 
The reason for this resides in the fact that the definition of the coupling is not as 
straightforward as for the model on fully connected graphs. In the latter scenario, the number of neighbors of a given node scales linearly with $N$; it thus suffices to set the $K/N$ to assure
that the coupling is an intensive quantity. 

The connectivity in real and synthetic networks, on the other hand, scales differently with the number of oscillators, making the definition of the coupling function to be not unique and, therefore, motivating the formulation of the equations of motion as Eq.~\ref{eq:KM_net}. Nevertheless, the lack of an appropriate normalization has several major consequences to the collective dynamics of Kuramoto oscillators: (i) the vanishing character of $K_c$ in the thermodynamic of limit for SF networks, as seen in the previous section; (ii) the difficulty in comparing the dynamics of networks with different connectivity patterns~\cite{arenas2008synchronization}, and (iii) the second term in the r.h.s. of Eq.~\ref{eq:KM_net} diverges in the thermodynamic limit for networks in which the maximum degree is not bounded when $N\rightarrow\infty$. In this section, we compare the impact of different prescriptions for the coupling function in large heterogeneous networks in the light of the latter points.

\begin{figure}[!tpb]
 \centerline{\includegraphics[width=0.9\columnwidth]{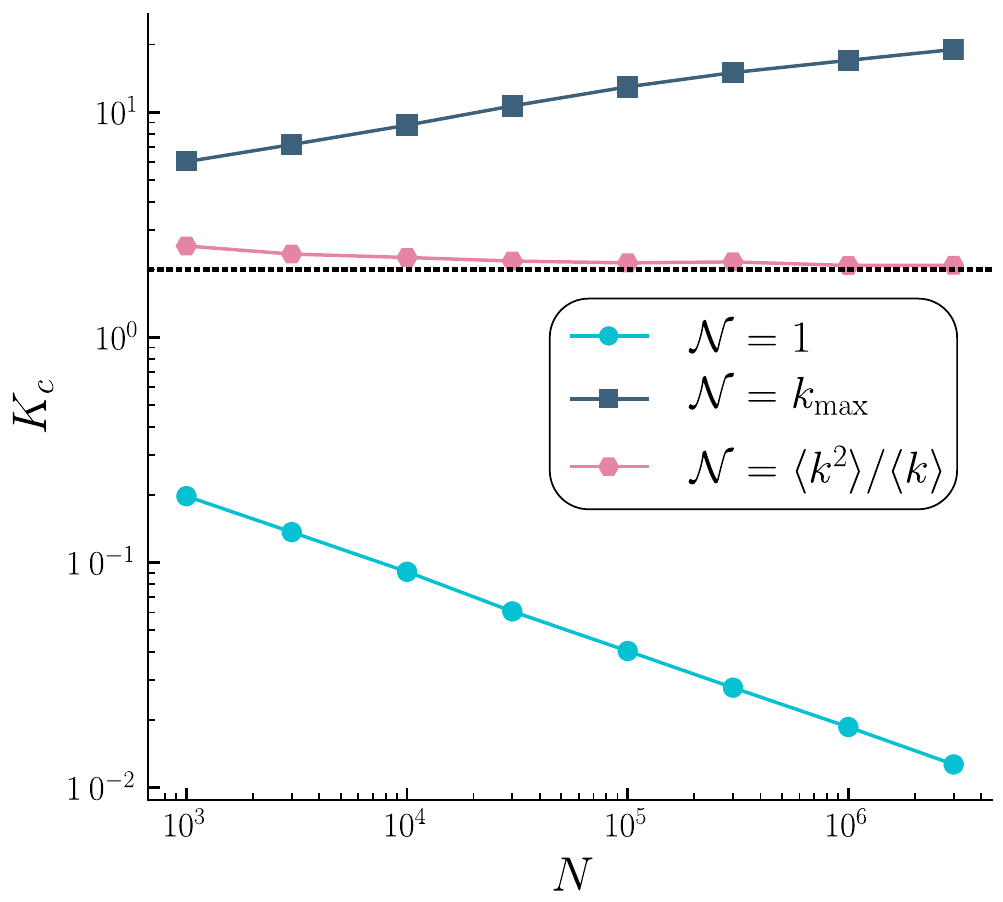}}
  \caption{(Color online) Numerical calculation of critical coupling $K_c$ as a function of number of oscillators $N$ of SF networks with $\gamma = 2.25$ under different normalizations $\mathcal{N}$. Dashed line marks the result $K_c = 2/[\pi g(0)]$. Natural frequencies distributed according to $g(\omega) = 1/\pi (\omega^2 + 1)$. Each point is an average over 100 network realizations. Error bars are smaller than symbols.}
  \label{fig:normalization}
\end{figure}

Let us now consider the phase equations defined as 
\begin{equation}
\dot{\theta}_i = \omega_i + \frac{K}{\mathcal{N}_i} \sum_{j=1}^N A_{ij} \sin(\theta_j - \theta_i),
\label{eq:KM_normalized}
\end{equation}
where $\mathcal{N}_i$ is the normalization constant of node $i$. Reasonable choices for $\mathcal{N}_i$ 
would then be quantities that are related to the network topology. One of these prescriptions discussed
in previous works is $\mathcal{N}_i = k_{\max}$ $\forall i$~\cite{arenas2008synchronization,rodrigues2016kuramoto}. It makes the summation to be an intensive quantity, since it prevents this term from diverging in highly heterogeneous networks. However, by repeating the analysis of the previous section for SF networks with $\mathcal{N} = k_{\max}$, we observe that the new normalization yields a critical coupling that depends on the system's size (Fig.~\ref{fig:normalization}). This result is easily understood by noticing that, under the HMF approximation and considering $P(k) \sim k^{-\gamma}$, $K_c$ is rescaled to
\begin{equation}
K_c = \frac{2}{\pi g(0)} \frac{\langle k \rangle }{\langle k^2 \rangle } k_{\max} \sim N^{\frac{\gamma}{2} - 1},
\label{eq:Kc_Kmax}
\end{equation}
explaning why the onset of synchronization increases in this case. Curiously, this conclusion is not evident from early works~\cite{arenas2008synchronization,rodrigues2016kuramoto}. Phenomenologically, the previous result can be understood by noticing that this normalization also makes the coupling $\frac{K}{\mathcal{N}_i}\rightarrow 0$ when $N\rightarrow\infty$ for all nodes with bounded connectivity in the thermodynamic limit. Thus, as these degree-bounded nodes are effectively decoupled of the hubs, one should expect $K_c\rightarrow\infty$ when $N\rightarrow\infty$.

\begin{figure}[!t]
 \centerline{\includegraphics[width=0.9\columnwidth]{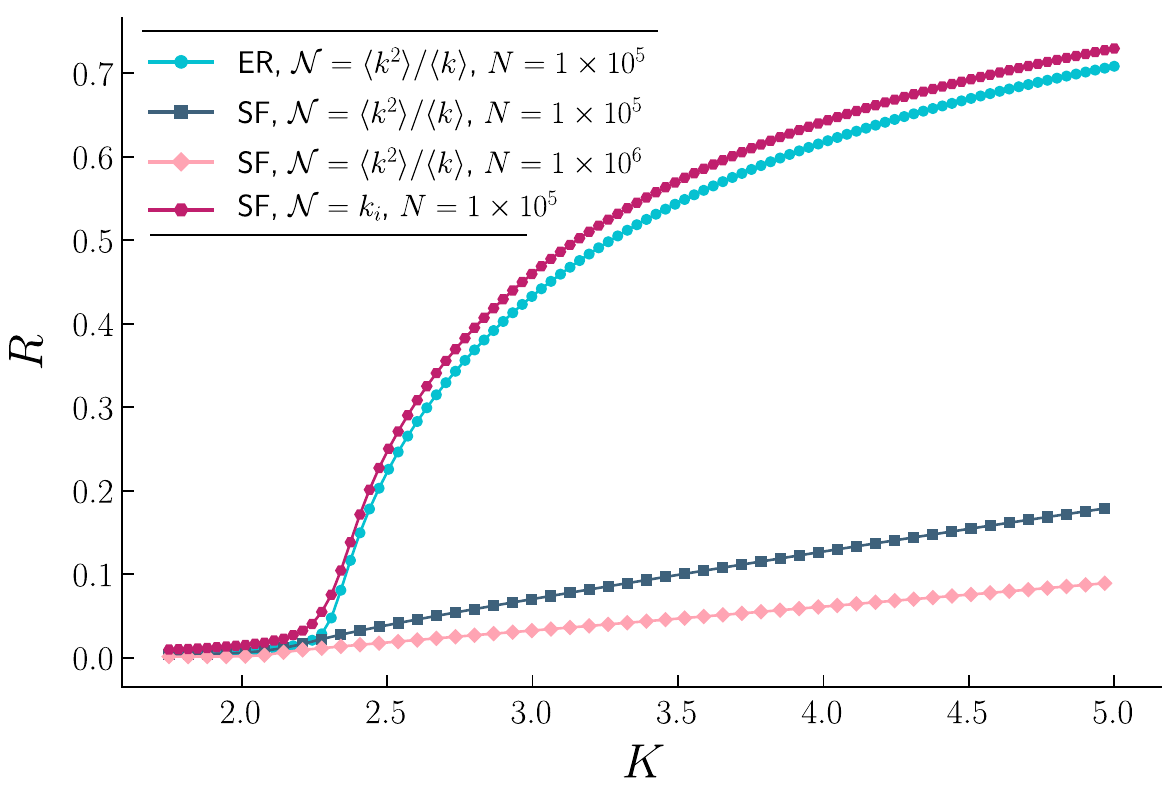}}
  \caption{(Color online) Synchronization curves considering different normalizations and network topologies. Natural frequencies distributed according to $g(\omega) = 1/\pi (\omega^2 + 1)$. SF networks considered in this figure have $\gamma = 2.25$ and $k_{\min}=5$. ER were generated with the same average degree as the SF networks with $N = 1\times 10^5$. Each point is an average over 100 network realizations. Error bars are smaller than symbols.}
  \label{fig:compar_ER}
\end{figure}

On the other hand, if a $K_c$ that is independent of the system's size is sought, then a natural choice would be
to rescale the coupling according to $\mathcal{N} = \langle k^2 \rangle / \langle k \rangle$.
Indeed, observing the corresponding result in Fig.~\ref{fig:normalization}, it looks like as if the problem 
of finding the appropriate normalization has been solved: as $N \rightarrow \infty$, $K_c$ converges to $2/\pi g(0)$, which is the same value encountered for the fully 
connected graph. Nevertheless, while this choice leads to a finite onset of synchronization in the 
thermodynamic limit -- and moreover sets the same $K_c$ for all heterogeneous networks -- , it imposes 
a vanishing coupling strength to low connected nodes. In other words, for infinitely large networks,  
such nodes will require an infinite $K$ in order to lock in synchrony with mean-field. This effect is evident in Fig.~\ref{fig:compar_ER}, 
where we see that even though SF networks with $\mathcal{N} = \langle k^2 \rangle / \langle k \rangle$ have similar $K_c$, the level of synchronization for $K > K_c$ decreases as $N$ gets larger. The solution for the problems of having a vanishing critical coupling and a diverging normalization for poorly connected nodes seems to be 
the choice $\mathcal{N}_i = k_i$. However, this comes with the price of washing out from the dynamics effects
that are intrinsic to the network topology, since the normalization acts as an average over the contribution of the nearest neighbors~\cite{arenas2008synchronization,rodrigues2016kuramoto}. For instance, as seen in Fig.~\ref{fig:compar_ER}, the synchronization curves of large ER and SF networks become qualitatively equivalent under $\mathcal{N}_i = k_i$.  What would then be an appropriate normalization for the coupling function for the Kuramoto dynamics on networks? It turns out that, if differences between the 
network structures must be highlighted, the most natural choice is the classical normalization $\mathcal{N} =1$, at the expense of having a vanishing $K_c$ for large networks with diverging $\langle k^2 \rangle$.
  
\section{Conclusion}
\label{sec:conclusion}

In this paper, we have analyzed the onset of synchronization of Kuramoto phase oscillators
in scale-free networks. First, we revisited a long-standing problem 
about the dynamics of Kuramoto oscillators coupled in heterogeneous topologies, namely, whether there is a nonzero critical coupling for the onset of synchronization in SF networks.  The debate around this question arose already in the early days of network science and, although there has been a substantial amount of work on network dynamics, this question has been seldom addressed in the last years.  
For SF networks with $\gamma < 3$, our extensive simulations showed that QMF and HMF solutions turned out to be equivalent in estimating the critical coupling strength. Specifically, both theories predicted a vanishing critical value for the onset of synchronization. For $\gamma > 3$, on the other hand, the HMF correctly predicted the finite threshold in the thermodynamic limit, whereas the QMF erroneously estimated a decaying critical coupling. 

We pointed out that this is a noticeable difference between the critical properties of synchronization of phase oscillators and the SIS dynamics. In particular, concerning the latter dynamics, experimental evidence reveals~\cite{ferreira2012epidemic,mata2013pair} that critical thresholds of SF networks with $\gamma > 3$ decrease
as $N \rightarrow \infty$, although with a different scaling as yielded by QMF. Nevertheless, in this regime of $\gamma$, the latter approximation estimates correctly secondary peaks in susceptibility curves associated to localization effects due to the epidemic activation of the largest hub in the network -- a phenomenon for which we have not observed a counterpart in the synchronization dynamics of large SF networks. Synchronization thresholds, on the other hand, present the same behavior as
observed in most dynamical processes that exhibit a phase transition from active to inactive states, such as 
contact process and SIRS model \cite{sander,Cota2018}. In fact, the phase transition observed in the Kuramoto model is a 
standard phase transition associated to a collective phenomenon (i.e., the activation of the entire network), 
whereas the phase transition in the SIS model is related to a mutual reinfection of hubs. Therefore, although synchronization and SIS epidemic thresholds behave similarly in SF networks with $ \gamma \leq 3$,
fundamental differences between the critical properties of these dynamics emerge for $\gamma > 3$. 
Future investigations should test if other types of localization effects and multiple transitions~\cite{arruda2017disease,colomer2014double} can be detectable in ensembles of phase oscillators.

In addition, we have also discussed the influence of different normalization
choices in the long term dynamics of large networks. We pointed out that choices previously considered
to be suitable for the dynamics of highly heterogeneous networks actually have
major drawbacks. Regarding the normalization by the maximum degree, while it prevents the hub's interaction 
to diverge, it yields a critical coupling that grows with the network's size -- a fact that remained unnoticed
in previous works. Normalizing the coupling function by the ratio $\langle k^2 \rangle/ \langle k \rangle$ circumvents the inconvenient of a size-dependent threshold. However, as for the case $\mathcal{N}_i=k_{\max}$, it ends up 
establishing a divergent normalization for low connected nodes, which requires them to have an infinite coupling strength to lock with the mean-field in the thermodynamic limit. The alternative then to these drawbacks is the
choice $\mathcal{N}_i = k_i$, which, as discussed here and in previous texts~\cite{arenas2008synchronization,rodrigues2016kuramoto}, removes from the dynamics the contribution from network topology, making networks with significantly different structures to 
exhibit similar synchronous dynamics. While this prescription could be appropriate 
in cases in which the focus of the analysis is not on the role played by the network topology in the dynamics (e.g.~\cite{jorg2014synchronization}), it seems counterintuitive that large heterogeneous networks should synchronize similarly as homogeneous ones. This scenario, therefore, points back to the conclusion that the most 
natural choice for the interaction between oscillators is the classical unnormalized couplings.

\section*{Acknowledgments}

TP acknowledges FAPESP (Grants No. 2016/23827-6 and 2018/15589-3). BM thanks CAPES for financial support.  
FAR acknowledges the Leverhulme Trust,
CNPq (Grant No. 305940/2010-4) and FAPESP (Grants
No. 2016/25682-5 and grants 2013/07375-0) for the financial
support given to this research. AM acknowledges FAPEMIG and CNPq (Grant No. 423185/2018-7). YM acknowledges partial support from the Government of Arag\'on, Spain through grant E36-17R (FENOL), by MINECO and FEDER funds (FIS2017-87519-P) and by Intesa Sanpaolo Innovation Center. This research was carried out using the computational resources of the Center for Mathematical Sciences Applied to Industry (CeMEAI) funded by FAPESP (grant 2013/07375-0). The funders had no role in study design, data collection and analysis, or preparation of the manuscript. 

\bibliography{bibliography}

\begin{thebibliography}{35}%
\makeatletter
\providecommand \@ifxundefined [1]{%
 \@ifx{#1\undefined}
}%
\providecommand \@ifnum [1]{%
 \ifnum #1\expandafter \@firstoftwo
 \else \expandafter \@secondoftwo
 \fi
}%
\providecommand \@ifx [1]{%
 \ifx #1\expandafter \@firstoftwo
 \else \expandafter \@secondoftwo
 \fi
}%
\providecommand \natexlab [1]{#1}%
\providecommand \enquote  [1]{``#1''}%
\providecommand \bibnamefont  [1]{#1}%
\providecommand \bibfnamefont [1]{#1}%
\providecommand \citenamefont [1]{#1}%
\providecommand \href@noop [0]{\@secondoftwo}%
\providecommand \href [0]{\begingroup \@sanitize@url \@href}%
\providecommand \@href[1]{\@@startlink{#1}\@@href}%
\providecommand \@@href[1]{\endgroup#1\@@endlink}%
\providecommand \@sanitize@url [0]{\catcode `\\12\catcode `\$12\catcode
  `\&12\catcode `\#12\catcode `\^12\catcode `\_12\catcode `\%12\relax}%
\providecommand \@@startlink[1]{}%
\providecommand \@@endlink[0]{}%
\providecommand \url  [0]{\begingroup\@sanitize@url \@url }%
\providecommand \@url [1]{\endgroup\@href {#1}{\urlprefix }}%
\providecommand \urlprefix  [0]{URL }%
\providecommand \Eprint [0]{\href }%
\providecommand \doibase [0]{http://dx.doi.org/}%
\providecommand \selectlanguage [0]{\@gobble}%
\providecommand \bibinfo  [0]{\@secondoftwo}%
\providecommand \bibfield  [0]{\@secondoftwo}%
\providecommand \translation [1]{[#1]}%
\providecommand \BibitemOpen [0]{}%
\providecommand \bibitemStop [0]{}%
\providecommand \bibitemNoStop [0]{.\EOS\space}%
\providecommand \EOS [0]{\spacefactor3000\relax}%
\providecommand \BibitemShut  [1]{\csname bibitem#1\endcsname}%
\let\auto@bib@innerbib\@empty
\bibitem [{\citenamefont {Pikovsky}\ \emph {et~al.}(2003)\citenamefont
  {Pikovsky}, \citenamefont {Rosenblum},\ and\ \citenamefont
  {Kurths}}]{pikovsky2003synchronization}%
  \BibitemOpen
  \bibfield  {author} {\bibinfo {author} {\bibfnamefont {Arkady}\ \bibnamefont
  {Pikovsky}}, \bibinfo {author} {\bibfnamefont {Michael}\ \bibnamefont
  {Rosenblum}}, \ and\ \bibinfo {author} {\bibfnamefont {J{\"u}rgen}\
  \bibnamefont {Kurths}},\ }\href@noop {} {\emph {\bibinfo {title}
  {Synchronization: a universal concept in nonlinear sciences}}},\
  Vol.~\bibinfo {volume} {12}\ (\bibinfo  {publisher} {Cambridge university
  press},\ \bibinfo {year} {2003})\BibitemShut {NoStop}%
\bibitem [{\citenamefont {Acebr{\'o}n}\ \emph {et~al.}(2005)\citenamefont
  {Acebr{\'o}n}, \citenamefont {Bonilla}, \citenamefont {Vicente},
  \citenamefont {Ritort},\ and\ \citenamefont {Spigler}}]{acebron2005kuramoto}%
  \BibitemOpen
  \bibfield  {author} {\bibinfo {author} {\bibfnamefont {Juan~A}\ \bibnamefont
  {Acebr{\'o}n}}, \bibinfo {author} {\bibfnamefont {Luis~L}\ \bibnamefont
  {Bonilla}}, \bibinfo {author} {\bibfnamefont {Conrad J~P{\'e}rez}\
  \bibnamefont {Vicente}}, \bibinfo {author} {\bibfnamefont {F{\'e}lix}\
  \bibnamefont {Ritort}}, \ and\ \bibinfo {author} {\bibfnamefont {Renato}\
  \bibnamefont {Spigler}},\ }\bibfield  {title} {\enquote {\bibinfo {title}
  {The {K}uramoto model: A simple paradigm for synchronization phenomena},}\
  }\href@noop {} {\bibfield  {journal} {\bibinfo  {journal} {Reviews of modern
  physics}\ }\textbf {\bibinfo {volume} {77}},\ \bibinfo {pages} {137}
  (\bibinfo {year} {2005})}\BibitemShut {NoStop}%
\bibitem [{\citenamefont {Arenas}\ \emph {et~al.}(2008)\citenamefont {Arenas},
  \citenamefont {D{\'\i}az-Guilera}, \citenamefont {Kurths}, \citenamefont
  {Moreno},\ and\ \citenamefont {Zhou}}]{arenas2008synchronization}%
  \BibitemOpen
  \bibfield  {author} {\bibinfo {author} {\bibfnamefont {Alex}\ \bibnamefont
  {Arenas}}, \bibinfo {author} {\bibfnamefont {Albert}\ \bibnamefont
  {D{\'\i}az-Guilera}}, \bibinfo {author} {\bibfnamefont {Jurgen}\ \bibnamefont
  {Kurths}}, \bibinfo {author} {\bibfnamefont {Yamir}\ \bibnamefont {Moreno}},
  \ and\ \bibinfo {author} {\bibfnamefont {Changsong}\ \bibnamefont {Zhou}},\
  }\bibfield  {title} {\enquote {\bibinfo {title} {Synchronization in complex
  networks},}\ }\href@noop {} {\bibfield  {journal} {\bibinfo  {journal}
  {Physics {R}eports}\ }\textbf {\bibinfo {volume} {469}},\ \bibinfo {pages}
  {93--153} (\bibinfo {year} {2008})}\BibitemShut {NoStop}%
\bibitem [{\citenamefont {Rodrigues}\ \emph {et~al.}(2016)\citenamefont
  {Rodrigues}, \citenamefont {Peron}, \citenamefont {Ji},\ and\ \citenamefont
  {Kurths}}]{rodrigues2016kuramoto}%
  \BibitemOpen
  \bibfield  {author} {\bibinfo {author} {\bibfnamefont {Francisco~A}\
  \bibnamefont {Rodrigues}}, \bibinfo {author} {\bibfnamefont {Thomas K~DM}\
  \bibnamefont {Peron}}, \bibinfo {author} {\bibfnamefont {Peng}\ \bibnamefont
  {Ji}}, \ and\ \bibinfo {author} {\bibfnamefont {J{\"u}rgen}\ \bibnamefont
  {Kurths}},\ }\bibfield  {title} {\enquote {\bibinfo {title} {The {K}uramoto
  model in complex networks},}\ }\href@noop {} {\bibfield  {journal} {\bibinfo
  {journal} {Physics {R}eports}\ }\textbf {\bibinfo {volume} {610}},\ \bibinfo
  {pages} {1--98} (\bibinfo {year} {2016})}\BibitemShut {NoStop}%
\bibitem [{\citenamefont {Moreno}\ and\ \citenamefont
  {Pacheco}(2004)}]{moreno2004synchronization}%
  \BibitemOpen
  \bibfield  {author} {\bibinfo {author} {\bibfnamefont {Yamir}\ \bibnamefont
  {Moreno}}\ and\ \bibinfo {author} {\bibfnamefont {Amalio~F}\ \bibnamefont
  {Pacheco}},\ }\bibfield  {title} {\enquote {\bibinfo {title} {Synchronization
  of {K}uramoto oscillators in scale-free networks},}\ }\href@noop {}
  {\bibfield  {journal} {\bibinfo  {journal} {EPL (Europhysics Letters)}\
  }\textbf {\bibinfo {volume} {68}},\ \bibinfo {pages} {603} (\bibinfo {year}
  {2004})}\BibitemShut {NoStop}%
\bibitem [{\citenamefont {Ichinomiya}(2004)}]{ichinomiya2004frequency}%
  \BibitemOpen
  \bibfield  {author} {\bibinfo {author} {\bibfnamefont {Takashi}\ \bibnamefont
  {Ichinomiya}},\ }\bibfield  {title} {\enquote {\bibinfo {title} {Frequency
  synchronization in a random oscillator network},}\ }\href@noop {} {\bibfield
  {journal} {\bibinfo  {journal} {Physical Review E}\ }\textbf {\bibinfo
  {volume} {70}},\ \bibinfo {pages} {026116} (\bibinfo {year}
  {2004})}\BibitemShut {NoStop}%
\bibitem [{\citenamefont {Lee}(2005)}]{lee2005synchronization}%
  \BibitemOpen
  \bibfield  {author} {\bibinfo {author} {\bibfnamefont {Deok-Sun}\
  \bibnamefont {Lee}},\ }\bibfield  {title} {\enquote {\bibinfo {title}
  {Synchronization transition in scale-free networks: Clusters of synchrony},}\
  }\href@noop {} {\bibfield  {journal} {\bibinfo  {journal} {Physical Review
  E}\ }\textbf {\bibinfo {volume} {72}},\ \bibinfo {pages} {026208} (\bibinfo
  {year} {2005})}\BibitemShut {NoStop}%
\bibitem [{\citenamefont {Boccaletti}\ \emph {et~al.}(2006)\citenamefont
  {Boccaletti}, \citenamefont {Latora}, \citenamefont {Moreno}, \citenamefont
  {Chavez},\ and\ \citenamefont {Hwang}}]{boccaletti2006complex}%
  \BibitemOpen
  \bibfield  {author} {\bibinfo {author} {\bibfnamefont {Stefano}\ \bibnamefont
  {Boccaletti}}, \bibinfo {author} {\bibfnamefont {Vito}\ \bibnamefont
  {Latora}}, \bibinfo {author} {\bibfnamefont {Yamir}\ \bibnamefont {Moreno}},
  \bibinfo {author} {\bibfnamefont {Martin}\ \bibnamefont {Chavez}}, \ and\
  \bibinfo {author} {\bibfnamefont {D-U}\ \bibnamefont {Hwang}},\ }\bibfield
  {title} {\enquote {\bibinfo {title} {Complex networks: Structure and
  dynamics},}\ }\href@noop {} {\bibfield  {journal} {\bibinfo  {journal}
  {Physics Reports}\ }\textbf {\bibinfo {volume} {424}},\ \bibinfo {pages}
  {175--308} (\bibinfo {year} {2006})}\BibitemShut {NoStop}%
\bibitem [{\citenamefont {Restrepo}\ \emph {et~al.}(2005)\citenamefont
  {Restrepo}, \citenamefont {Ott},\ and\ \citenamefont
  {Hunt}}]{restrepo2005onset}%
  \BibitemOpen
  \bibfield  {author} {\bibinfo {author} {\bibfnamefont {Juan~G}\ \bibnamefont
  {Restrepo}}, \bibinfo {author} {\bibfnamefont {Edward}\ \bibnamefont {Ott}},
  \ and\ \bibinfo {author} {\bibfnamefont {Brian~R}\ \bibnamefont {Hunt}},\
  }\bibfield  {title} {\enquote {\bibinfo {title} {Onset of synchronization in
  large networks of coupled oscillators},}\ }\href@noop {} {\bibfield
  {journal} {\bibinfo  {journal} {Physical Review E}\ }\textbf {\bibinfo
  {volume} {71}},\ \bibinfo {pages} {036151} (\bibinfo {year}
  {2005})}\BibitemShut {NoStop}%
\bibitem [{\citenamefont {Dorogovtsev}(2010)}]{dorogovtsev2010lectures}%
  \BibitemOpen
  \bibfield  {author} {\bibinfo {author} {\bibfnamefont {Sergei~N}\
  \bibnamefont {Dorogovtsev}},\ }\href@noop {} {\emph {\bibinfo {title}
  {Lectures on complex networks}}},\ Vol.~\bibinfo {volume} {24}\ (\bibinfo
  {publisher} {Oxford University Press Oxford},\ \bibinfo {year}
  {2010})\BibitemShut {NoStop}%
\bibitem [{\citenamefont {Hong}\ \emph {et~al.}(2007)\citenamefont {Hong},
  \citenamefont {Park},\ and\ \citenamefont {Tang}}]{hong2007finite}%
  \BibitemOpen
  \bibfield  {author} {\bibinfo {author} {\bibfnamefont {Hyunsuk}\ \bibnamefont
  {Hong}}, \bibinfo {author} {\bibfnamefont {Hyunggyu}\ \bibnamefont {Park}}, \
  and\ \bibinfo {author} {\bibfnamefont {Lei-Han}\ \bibnamefont {Tang}},\
  }\bibfield  {title} {\enquote {\bibinfo {title} {Finite-size scaling of
  synchronized oscillation on complex networks},}\ }\href@noop {} {\bibfield
  {journal} {\bibinfo  {journal} {Physical Review E}\ }\textbf {\bibinfo
  {volume} {76}},\ \bibinfo {pages} {066104} (\bibinfo {year}
  {2007})}\BibitemShut {NoStop}%
\bibitem [{\citenamefont {Hong}\ \emph {et~al.}(2013)\citenamefont {Hong},
  \citenamefont {Um},\ and\ \citenamefont {Park}}]{hong2013link}%
  \BibitemOpen
  \bibfield  {author} {\bibinfo {author} {\bibfnamefont {Hyunsuk}\ \bibnamefont
  {Hong}}, \bibinfo {author} {\bibfnamefont {Jaegon}\ \bibnamefont {Um}}, \
  and\ \bibinfo {author} {\bibfnamefont {Hyunggyu}\ \bibnamefont {Park}},\
  }\bibfield  {title} {\enquote {\bibinfo {title} {Link-disorder fluctuation
  effects on synchronization in random networks},}\ }\href@noop {} {\bibfield
  {journal} {\bibinfo  {journal} {Physical Review E}\ }\textbf {\bibinfo
  {volume} {87}},\ \bibinfo {pages} {042105} (\bibinfo {year}
  {2013})}\BibitemShut {NoStop}%
\bibitem [{\citenamefont {Um}\ \emph {et~al.}(2014)\citenamefont {Um},
  \citenamefont {Hong},\ and\ \citenamefont {Park}}]{um2014nature}%
  \BibitemOpen
  \bibfield  {author} {\bibinfo {author} {\bibfnamefont {Jaegon}\ \bibnamefont
  {Um}}, \bibinfo {author} {\bibfnamefont {Hyunsuk}\ \bibnamefont {Hong}}, \
  and\ \bibinfo {author} {\bibfnamefont {Hyunggyu}\ \bibnamefont {Park}},\
  }\bibfield  {title} {\enquote {\bibinfo {title} {Nature of synchronization
  transitions in random networks of coupled oscillators},}\ }\href@noop {}
  {\bibfield  {journal} {\bibinfo  {journal} {Physical Review E}\ }\textbf
  {\bibinfo {volume} {89}},\ \bibinfo {pages} {012810} (\bibinfo {year}
  {2014})}\BibitemShut {NoStop}%
\bibitem [{\citenamefont {Juh{\'a}sz}\ \emph {et~al.}(2019)\citenamefont
  {Juh{\'a}sz}, \citenamefont {Kelling},\ and\ \citenamefont
  {Odor}}]{juhasz2019critical}%
  \BibitemOpen
  \bibfield  {author} {\bibinfo {author} {\bibfnamefont {R{\'o}bert}\
  \bibnamefont {Juh{\'a}sz}}, \bibinfo {author} {\bibfnamefont {Jeffrey}\
  \bibnamefont {Kelling}}, \ and\ \bibinfo {author} {\bibfnamefont {G{\'e}za}\
  \bibnamefont {Odor}},\ }\bibfield  {title} {\enquote {\bibinfo {title}
  {Critical dynamics of the {K}uramoto model on sparse random networks},}\
  }\href@noop {} {\bibfield  {journal} {\bibinfo  {journal} {arXiv preprint
  arXiv:1902.10422}\ } (\bibinfo {year} {2019})}\BibitemShut {NoStop}%
\bibitem [{\citenamefont {Yook}\ and\ \citenamefont {Kim}(2018)}]{yook2018two}%
  \BibitemOpen
  \bibfield  {author} {\bibinfo {author} {\bibfnamefont {Soon-Hyung}\
  \bibnamefont {Yook}}\ and\ \bibinfo {author} {\bibfnamefont {Yup}\
  \bibnamefont {Kim}},\ }\bibfield  {title} {\enquote {\bibinfo {title} {Two
  order parameters for the {K}uramoto model on complex networks},}\ }\href@noop
  {} {\bibfield  {journal} {\bibinfo  {journal} {Physical Review E}\ }\textbf
  {\bibinfo {volume} {97}},\ \bibinfo {pages} {042317} (\bibinfo {year}
  {2018})}\BibitemShut {NoStop}%
\bibitem [{\citenamefont {Ferreira}\ \emph {et~al.}(2012)\citenamefont
  {Ferreira}, \citenamefont {Castellano},\ and\ \citenamefont
  {Pastor-Satorras}}]{ferreira2012epidemic}%
  \BibitemOpen
  \bibfield  {author} {\bibinfo {author} {\bibfnamefont {Silvio~C}\
  \bibnamefont {Ferreira}}, \bibinfo {author} {\bibfnamefont {Claudio}\
  \bibnamefont {Castellano}}, \ and\ \bibinfo {author} {\bibfnamefont
  {Romualdo}\ \bibnamefont {Pastor-Satorras}},\ }\bibfield  {title} {\enquote
  {\bibinfo {title} {Epidemic thresholds of the
  susceptible-infected-susceptible model on networks: A comparison of numerical
  and theoretical results},}\ }\href@noop {} {\bibfield  {journal} {\bibinfo
  {journal} {Physical Review E}\ }\textbf {\bibinfo {volume} {86}},\ \bibinfo
  {pages} {041125} (\bibinfo {year} {2012})}\BibitemShut {NoStop}%
\bibitem [{\citenamefont {Mata}\ and\ \citenamefont
  {Ferreira}(2013)}]{mata2013pair}%
  \BibitemOpen
  \bibfield  {author} {\bibinfo {author} {\bibfnamefont {Ang{\'e}lica~S}\
  \bibnamefont {Mata}}\ and\ \bibinfo {author} {\bibfnamefont {Silvio~C}\
  \bibnamefont {Ferreira}},\ }\bibfield  {title} {\enquote {\bibinfo {title}
  {Pair quenched mean-field theory for the susceptible-infected-susceptible
  model on complex networks},}\ }\href@noop {} {\bibfield  {journal} {\bibinfo
  {journal} {EPL (Europhysics Letters)}\ }\textbf {\bibinfo {volume} {103}},\
  \bibinfo {pages} {48003} (\bibinfo {year} {2013})}\BibitemShut {NoStop}%
\bibitem [{\citenamefont {Castellano}\ and\ \citenamefont
  {Pastor-Satorras}(2010)}]{castellano2010thresholds}%
  \BibitemOpen
  \bibfield  {author} {\bibinfo {author} {\bibfnamefont {Claudio}\ \bibnamefont
  {Castellano}}\ and\ \bibinfo {author} {\bibfnamefont {Romualdo}\ \bibnamefont
  {Pastor-Satorras}},\ }\bibfield  {title} {\enquote {\bibinfo {title}
  {Thresholds for epidemic spreading in networks},}\ }\href@noop {} {\bibfield
  {journal} {\bibinfo  {journal} {Physical {R}eview {L}etters}\ }\textbf
  {\bibinfo {volume} {105}},\ \bibinfo {pages} {218701} (\bibinfo {year}
  {2010})}\BibitemShut {NoStop}%
\bibitem [{\citenamefont {Chung}\ \emph {et~al.}(2004)\citenamefont {Chung},
  \citenamefont {Lu},\ and\ \citenamefont {Vu}}]{chung2004spectra}%
  \BibitemOpen
  \bibfield  {author} {\bibinfo {author} {\bibfnamefont {Fan}\ \bibnamefont
  {Chung}}, \bibinfo {author} {\bibfnamefont {Linyuan}\ \bibnamefont {Lu}}, \
  and\ \bibinfo {author} {\bibfnamefont {Van}\ \bibnamefont {Vu}},\ }\bibfield
  {title} {\enquote {\bibinfo {title} {The spectra of random graphs with given
  expected degrees},}\ }\href@noop {} {\bibfield  {journal} {\bibinfo
  {journal} {Internet Mathematics}\ }\textbf {\bibinfo {volume} {1}},\ \bibinfo
  {pages} {257--275} (\bibinfo {year} {2004})}\BibitemShut {NoStop}%
\bibitem [{\citenamefont {Catanzaro}\ \emph {et~al.}(2005)\citenamefont
  {Catanzaro}, \citenamefont {Bogun{\'a}},\ and\ \citenamefont
  {Pastor-Satorras}}]{catanzaro2005generation}%
  \BibitemOpen
  \bibfield  {author} {\bibinfo {author} {\bibfnamefont {Michele}\ \bibnamefont
  {Catanzaro}}, \bibinfo {author} {\bibfnamefont {Mari{\'a}n}\ \bibnamefont
  {Bogun{\'a}}}, \ and\ \bibinfo {author} {\bibfnamefont {Romualdo}\
  \bibnamefont {Pastor-Satorras}},\ }\bibfield  {title} {\enquote {\bibinfo
  {title} {Generation of uncorrelated random scale-free networks},}\
  }\href@noop {} {\bibfield  {journal} {\bibinfo  {journal} {Physical Review
  E}\ }\textbf {\bibinfo {volume} {71}},\ \bibinfo {pages} {027103} (\bibinfo
  {year} {2005})}\BibitemShut {NoStop}%
\bibitem [{\citenamefont {Mata}\ and\ \citenamefont
  {Ferreira}(2015)}]{mata2015multiple}%
  \BibitemOpen
  \bibfield  {author} {\bibinfo {author} {\bibfnamefont {Ang{\'e}lica~S}\
  \bibnamefont {Mata}}\ and\ \bibinfo {author} {\bibfnamefont {Silvio~C}\
  \bibnamefont {Ferreira}},\ }\bibfield  {title} {\enquote {\bibinfo {title}
  {Multiple transitions of the susceptible-infected-susceptible epidemic model
  on complex networks},}\ }\href@noop {} {\bibfield  {journal} {\bibinfo
  {journal} {Physical Review E}\ }\textbf {\bibinfo {volume} {91}},\ \bibinfo
  {pages} {012816} (\bibinfo {year} {2015})}\BibitemShut {NoStop}%
\bibitem [{\citenamefont {Mata}\ \emph {et~al.}(2014)\citenamefont {Mata},
  \citenamefont {Ferreira},\ and\ \citenamefont
  {Ferreira}}]{mata2014heterogeneous}%
  \BibitemOpen
  \bibfield  {author} {\bibinfo {author} {\bibfnamefont {Ang{\'e}lica~S}\
  \bibnamefont {Mata}}, \bibinfo {author} {\bibfnamefont {Ronan~S}\
  \bibnamefont {Ferreira}}, \ and\ \bibinfo {author} {\bibfnamefont {Silvio~C}\
  \bibnamefont {Ferreira}},\ }\bibfield  {title} {\enquote {\bibinfo {title}
  {Heterogeneous pair-approximation for the contact process on complex
  networks},}\ }\href@noop {} {\bibfield  {journal} {\bibinfo  {journal} {New
  Journal of Physics}\ }\textbf {\bibinfo {volume} {16}},\ \bibinfo {pages}
  {053006} (\bibinfo {year} {2014})}\BibitemShut {NoStop}%
\bibitem [{\citenamefont {Harris}(1974)}]{harris74}%
  \BibitemOpen
  \bibfield  {author} {\bibinfo {author} {\bibfnamefont {T.~E.}\ \bibnamefont
  {Harris}},\ }\bibfield  {title} {\enquote {\bibinfo {title} {Contact
  interactions on a lattice},}\ }\href@noop {} {\bibfield  {journal} {\bibinfo
  {journal} {Ann. Prob.}\ }\textbf {\bibinfo {volume} {2}},\ \bibinfo {pages}
  {969--988} (\bibinfo {year} {1974})}\BibitemShut {NoStop}%
\bibitem [{\citenamefont {Diekmann}\ and\ \citenamefont
  {Heesterbeek}(2000)}]{epidemics}%
  \BibitemOpen
  \bibfield  {author} {\bibinfo {author} {\bibfnamefont {O.}~\bibnamefont
  {Diekmann}}\ and\ \bibinfo {author} {\bibfnamefont {J.A.P}\ \bibnamefont
  {Heesterbeek}},\ }\href@noop {} {\emph {\bibinfo {title} {Mathematical
  epidemiology of infectious diseases: model building, analysis and
  interpretation}}}\ (\bibinfo  {publisher} {John Wiley \& Sons},\ \bibinfo
  {address} {New York},\ \bibinfo {year} {2000})\BibitemShut {NoStop}%
\bibitem [{\citenamefont {Chatterjee}\ and\ \citenamefont
  {Durrett}(2009)}]{Chatterjee09}%
  \BibitemOpen
  \bibfield  {author} {\bibinfo {author} {\bibfnamefont {S.}~\bibnamefont
  {Chatterjee}}\ and\ \bibinfo {author} {\bibfnamefont {R.}~\bibnamefont
  {Durrett}},\ }\bibfield  {title} {\enquote {\bibinfo {title} {Contact
  processes on random graphs with power law degree distributions have critical
  value 0},}\ }\href@noop {} {\bibfield  {journal} {\bibinfo  {journal} {Ann.
  Probab.}\ }\textbf {\bibinfo {volume} {37}},\ \bibinfo {pages} {2332--2356}
  (\bibinfo {year} {2009})}\BibitemShut {NoStop}%
\bibitem [{\citenamefont {Bogu\~n\'a}\ \emph {et~al.}(2013)\citenamefont
  {Bogu\~n\'a}, \citenamefont {Castellano},\ and\ \citenamefont
  {Pastor-Satorras}}]{boguna2013nature}%
  \BibitemOpen
  \bibfield  {author} {\bibinfo {author} {\bibfnamefont {Marian}\ \bibnamefont
  {Bogu\~n\'a}}, \bibinfo {author} {\bibfnamefont {Claudio}\ \bibnamefont
  {Castellano}}, \ and\ \bibinfo {author} {\bibfnamefont {Romualdo}\
  \bibnamefont {Pastor-Satorras}},\ }\bibfield  {title} {\enquote {\bibinfo
  {title} {Nature of the epidemic threshold for the
  susceptible-infected-susceptible dynamics in networks},}\ }\href@noop {}
  {\bibfield  {journal} {\bibinfo  {journal} {Phys. Rev. Lett.}\ }\textbf
  {\bibinfo {volume} {111}},\ \bibinfo {pages} {068701} (\bibinfo {year}
  {2013})}\BibitemShut {NoStop}%
\bibitem [{\citenamefont {Cota}\ \emph {et~al.}(2016)\citenamefont {Cota},
  \citenamefont {Ferreira},\ and\ \citenamefont {\'Odor}}]{Cota}%
  \BibitemOpen
  \bibfield  {author} {\bibinfo {author} {\bibfnamefont {Wesley}\ \bibnamefont
  {Cota}}, \bibinfo {author} {\bibfnamefont {Silvio~C.}\ \bibnamefont
  {Ferreira}}, \ and\ \bibinfo {author} {\bibfnamefont {G\'eza}\ \bibnamefont
  {\'Odor}},\ }\bibfield  {title} {\enquote {\bibinfo {title} {Griffiths
  effects of the susceptible-infected-susceptible epidemic model on random
  power-law networks},}\ }\href {\doibase 10.1103/PhysRevE.93.032322}
  {\bibfield  {journal} {\bibinfo  {journal} {Phys. Rev. E}\ }\textbf {\bibinfo
  {volume} {93}},\ \bibinfo {pages} {032322} (\bibinfo {year}
  {2016})}\BibitemShut {NoStop}%
\bibitem [{\citenamefont {Ferreira}\ \emph {et~al.}(2016)\citenamefont
  {Ferreira}, \citenamefont {Sander},\ and\ \citenamefont
  {Pastor-Satorras}}]{sander}%
  \BibitemOpen
  \bibfield  {author} {\bibinfo {author} {\bibfnamefont {Silvio~C.}\
  \bibnamefont {Ferreira}}, \bibinfo {author} {\bibfnamefont {Renan~S.}\
  \bibnamefont {Sander}}, \ and\ \bibinfo {author} {\bibfnamefont {Romualdo}\
  \bibnamefont {Pastor-Satorras}},\ }\bibfield  {title} {\enquote {\bibinfo
  {title} {Collective versus hub activation of epidemic phases on networks},}\
  }\href {\doibase 10.1103/PhysRevE.93.032314} {\bibfield  {journal} {\bibinfo
  {journal} {Phys. Rev. E}\ }\textbf {\bibinfo {volume} {93}},\ \bibinfo
  {pages} {032314} (\bibinfo {year} {2016})}\BibitemShut {NoStop}%
\bibitem [{\citenamefont {Cota}\ \emph {et~al.}(2018)\citenamefont {Cota},
  \citenamefont {Mata},\ and\ \citenamefont {Ferreira}}]{Cota2018}%
  \BibitemOpen
  \bibfield  {author} {\bibinfo {author} {\bibfnamefont {Wesley}\ \bibnamefont
  {Cota}}, \bibinfo {author} {\bibfnamefont {Ang\'elica~S.}\ \bibnamefont
  {Mata}}, \ and\ \bibinfo {author} {\bibfnamefont {Silvio~C.}\ \bibnamefont
  {Ferreira}},\ }\bibfield  {title} {\enquote {\bibinfo {title} {Robustness and
  fragility of the susceptible-infected-susceptible epidemic models on complex
  networks},}\ }\href {\doibase 10.1103/PhysRevE.98.012310} {\bibfield
  {journal} {\bibinfo  {journal} {Phys. Rev. E}\ }\textbf {\bibinfo {volume}
  {98}},\ \bibinfo {pages} {012310} (\bibinfo {year} {2018})}\BibitemShut
  {NoStop}%
\bibitem [{\citenamefont {Anderson}\ and\ \citenamefont
  {May}(1992)}]{anderson}%
  \BibitemOpen
  \bibfield  {author} {\bibinfo {author} {\bibfnamefont {R.~M.}\ \bibnamefont
  {Anderson}}\ and\ \bibinfo {author} {\bibfnamefont {R.~M.}\ \bibnamefont
  {May}},\ }\href@noop {} {\emph {\bibinfo {title} {Infectious Diseases in
  Humans}}}\ (\bibinfo  {publisher} {Oxford University Press},\ \bibinfo
  {address} {Oxford},\ \bibinfo {year} {1992})\BibitemShut {NoStop}%
\bibitem [{\citenamefont {Castellano}\ and\ \citenamefont
  {Pastor-Satorras}(2006)}]{cp}%
  \BibitemOpen
  \bibfield  {author} {\bibinfo {author} {\bibfnamefont {Claudio}\ \bibnamefont
  {Castellano}}\ and\ \bibinfo {author} {\bibfnamefont {Romualdo}\ \bibnamefont
  {Pastor-Satorras}},\ }\bibfield  {title} {\enquote {\bibinfo {title}
  {Non-mean-field behavior of the contact process on scale-free networks},}\
  }\href {\doibase 10.1103/PhysRevLett.96.038701} {\bibfield  {journal}
  {\bibinfo  {journal} {Phys. Rev. Lett.}\ }\textbf {\bibinfo {volume} {96}},\
  \bibinfo {pages} {038701} (\bibinfo {year} {2006})}\BibitemShut {NoStop}%
\bibitem [{\citenamefont {Karsai}\ \emph {et~al.}(2006)\citenamefont {Karsai},
  \citenamefont {Juh\'asz},\ and\ \citenamefont {Igl\'oi}}]{KJI}%
  \BibitemOpen
  \bibfield  {author} {\bibinfo {author} {\bibfnamefont {M\'arton}\
  \bibnamefont {Karsai}}, \bibinfo {author} {\bibfnamefont {R\'obert}\
  \bibnamefont {Juh\'asz}}, \ and\ \bibinfo {author} {\bibfnamefont {Ferenc}\
  \bibnamefont {Igl\'oi}},\ }\bibfield  {title} {\enquote {\bibinfo {title}
  {Nonequilibrium phase transitions and finite-size scaling in weighted
  scale-free networks},}\ }\href {\doibase 10.1103/PhysRevE.73.036116}
  {\bibfield  {journal} {\bibinfo  {journal} {Phys. Rev. E}\ }\textbf {\bibinfo
  {volume} {73}},\ \bibinfo {pages} {036116} (\bibinfo {year}
  {2006})}\BibitemShut {NoStop}%
\bibitem [{\citenamefont {de~Arruda}\ \emph {et~al.}(2017)\citenamefont
  {de~Arruda}, \citenamefont {Cozzo}, \citenamefont {Peixoto}, \citenamefont
  {Rodrigues},\ and\ \citenamefont {Moreno}}]{arruda2017disease}%
  \BibitemOpen
  \bibfield  {author} {\bibinfo {author} {\bibfnamefont {Guilherme~Ferraz}\
  \bibnamefont {de~Arruda}}, \bibinfo {author} {\bibfnamefont {Emanuele}\
  \bibnamefont {Cozzo}}, \bibinfo {author} {\bibfnamefont {Tiago~P}\
  \bibnamefont {Peixoto}}, \bibinfo {author} {\bibfnamefont {Francisco~A}\
  \bibnamefont {Rodrigues}}, \ and\ \bibinfo {author} {\bibfnamefont {Yamir}\
  \bibnamefont {Moreno}},\ }\bibfield  {title} {\enquote {\bibinfo {title}
  {Disease localization in multilayer networks},}\ }\href@noop {} {\bibfield
  {journal} {\bibinfo  {journal} {Physical Review X}\ }\textbf {\bibinfo
  {volume} {7}},\ \bibinfo {pages} {011014} (\bibinfo {year}
  {2017})}\BibitemShut {NoStop}%
\bibitem [{\citenamefont {Colomer-de Sim{\'o}n}\ and\ \citenamefont
  {Bogu{\~n}{\'a}}(2014)}]{colomer2014double}%
  \BibitemOpen
  \bibfield  {author} {\bibinfo {author} {\bibfnamefont {Pol}\ \bibnamefont
  {Colomer-de Sim{\'o}n}}\ and\ \bibinfo {author} {\bibfnamefont {Mari{\'a}n}\
  \bibnamefont {Bogu{\~n}{\'a}}},\ }\bibfield  {title} {\enquote {\bibinfo
  {title} {Double percolation phase transition in clustered complex
  networks},}\ }\href@noop {} {\bibfield  {journal} {\bibinfo  {journal}
  {Physical Review X}\ }\textbf {\bibinfo {volume} {4}},\ \bibinfo {pages}
  {041020} (\bibinfo {year} {2014})}\BibitemShut {NoStop}%
\bibitem [{\citenamefont {J{\"o}rg}\ \emph {et~al.}(2014)\citenamefont
  {J{\"o}rg}, \citenamefont {Morelli}, \citenamefont {Ares},\ and\
  \citenamefont {J{\"u}licher}}]{jorg2014synchronization}%
  \BibitemOpen
  \bibfield  {author} {\bibinfo {author} {\bibfnamefont {David~J}\ \bibnamefont
  {J{\"o}rg}}, \bibinfo {author} {\bibfnamefont {Luis~G}\ \bibnamefont
  {Morelli}}, \bibinfo {author} {\bibfnamefont {Sa{\'u}l}\ \bibnamefont
  {Ares}}, \ and\ \bibinfo {author} {\bibfnamefont {Frank}\ \bibnamefont
  {J{\"u}licher}},\ }\bibfield  {title} {\enquote {\bibinfo {title}
  {Synchronization dynamics in the presence of coupling delays and phase
  shifts},}\ }\href@noop {} {\bibfield  {journal} {\bibinfo  {journal}
  {Physical {R}eview {L}etters}\ }\textbf {\bibinfo {volume} {112}},\ \bibinfo
  {pages} {174101} (\bibinfo {year} {2014})}\BibitemShut {NoStop}%
\end{thebibliography}%

\end{document}